# Autotuning by Changing Directives and Number of Threads in OpenMP using ppOpen-AT

Toma Sakurai
Graduate School of Informatics
Nagoya University
Japan

Satoshi Ohshima
Information Technology Center
Nagoya University
Japan

Takahiro Katagiri
Information Technology Center
Nagoya University
Japan
Contact: katagiri@cc.nagoya-u.ac.jp

Toru Nagai
Information Technology Center
Nagoya University
Japan

*Abstract*— Recently, computers have diversified architectures. To achieve high numerical calculation software performance, it is necessary to tune the software according to the target computer architecture. However, code optimization for each environment is difficult unless it is performed by a specialist who knows computer architectures well. By applying autotuning (AT), the tuning effort can be reduced. Optimized implementation by AT that enhances computer performance can be used even by non-experts. In this research, we propose a technique for AT for programs using open multi-processing (OpenMP). We propose an AT method using an AT language that changes the OpenMP optimized loop and dynamically changes the number of threads in OpenMP according to computational kernels. Performance evaluation was performed using the Fujitsu PRIMEHPC FX100, which is a K-computer type supercomputer installed at the Information Technology Center, Nagoya University. As a result, we found there was a performance increase of 1.801 times that of the original code in a plasma turbulence analysis.

## I. INTRODUCTION

In recent years, mainstream computer architectures have become multi-core central processing unit (CPU) architectures that include hierarchical memory structures and caches. The presence or absence of a graphics processing unit (GPU) varies. Software tuning is important to achieve high numerical computation software performance, but optimization for the computing environment requires specialized knowledge of hardware architectures. It is time-consuming work. In addition, programs tuned specifically for one environment may have poor performance in other environments, requiring further performance tuning. In addition, widely used compiler optimizations are often not useful for software developers for code optimization due to lack of knowledge of the compiler code.

Software autotuning (AT) [1], which is one of the code optimization technologies, automates performance tuning for programs. By using this technology, it is possible to achieve high performance with the same program in different environments. It is also possible to improve the performance portability of software. Therefore, in recent years, many numerical calculation software applications using AT have been developed [2][3][4]. In addition, AT frameworks have been proposed for creating numerically tuned software with AT. Typical examples include OpenTuner [5], Xevolver [6], and FIBER [7].

In this study we propose and evaluate the following three contents:
1. Propose an AT function for OpenMP directives for an AT language.
2. Propose an AT function to dynamically change the number of OpenMP threads.
3. Evaluate the efficiency for 1 and 2.

This paper is organized as follows. Section II explains the framework of the AT software and the AT languages for the paper. Section III proposes an AT function for exchanging OpenMP directives. Section IV also proposes an AT function to dynamically change the number of OpenMP threads. Section V evaluates the total efficiency for the two new functions. Section VI summarizes related work and finally, Section VII describes the conclusion to this paper.

## II. FIBER FRAMEWORK AND AT LANGAGE PPOPEN-AT

### A. FIBER

FIBER (framework of installation, before execution, and run-time optimization layers) [7] is an AT framework for use in numerical calculation software. This framework is a software configuration method that can perform AT at the following three time points (layers).
1. Installation: optimization hierarchy when installing a library.
2. Before execution: optimization hierarchy whose parameters (problem size, the number of processes for message passing interface (MPI), and number of OpenMP threads) have been determined by the user.
3. Run-time: optimization hierarchy when a library is executed.

With the above hierarchy, it is possible to increase the number of applications to which AT is applied and improve the accuracy of parameter estimation.

FIBER supports the following two functions.
1. A function that generates code that performs AT, parameterizes code, and registers parameters by giving instructions to the user's program using a dedicated language.
2. Parameter optimization in each hierarchy.

AT in FIBER is determined by the basic parameter set, the performance parameter set, and the cost definition function. The basic parameter set (BP) is a set of parameters related to basic information, such as matrix size, and the computer environment, such as number of processors and computer configuration method, when performing numerical calculations. The performance parameter set (PP) is a set of parameters that determine the program performance when fixing the BP. The cost definition function is a function that finds the cost determined by PP at each time point when BP is fixed. This cost can be considered as program execution time, memory usage, and power consumption.

AT in FIBER is defined as the process of finding the performance parameter set that minimizes the cost definition function when the basic parameter set BP determined at each time point is fixed.

### B. ppOpen-AT

ppOpen-AT [8] is an AT language developed as an AT mechanism for the next generation of science and technology application development. Its execution environment is ppOpen-HPC [9]. Designed to improve the developmental efficiency of parallel numerical computation libraries, it inherits the functions of the AT directive-based language ABCLibScript [10]. ABCLibScript supports the addition of an AT function using FIBER methods.

ppOpen-AT adds AT functions using FIBER to Fortran90 and C programs. Since descriptions about AT are dedicated directives, the AT software can be efficiently developed compared to development in environments without ppOpen-AT. The ppOpen-AT preprocessor interprets directives and generates code for tuning candidates and programs that include an AT function that searches for the optimal code.

Functions of AT created by ppOpen-AT are generated by rewriting the original program. The newly generated code includes tuning candidates before execution by end-users. Therefore, end-users do not have to generate their code at run-time in their own environments [11]. This method is a light load AT, without dynamic code generation. For example, it works well in the supercomputer environment because the log-in node has no load for AT, such as the code generation load. However, the amount of code may increase from the original code, since all tuning candidates are generated in advance and are included in the original program without AT functions. In order to prevent code expansion, software developers must target processes so that the number of tuning candidates are limited [12].

ppOpen-AT provides loop unrolling, code selection, loop split (split), loop fusion (collapse), and reordering of sentences as generated code. The generated code is included in the original program without AT functions as libraries for the candidates.

## III. A NEW AT FUNCTION: OPTIMIZED LOOP CHANGE UISNG OPENMP

### A. Proposed Method

We propose an AT function for loop transformation that changes the target parallelization loop. This can be implemented as a new function of ppOpen-AT. Our target is multiple nested loops being parallelized by OpenMP.
The following specifies the proposed method, a new directive created by ppOpen-AT to the optimized loop. The AT region is specified with region start – region end.

**!oat$ install Exchange (Number of loop depth, Number of loop depth, ...) region start**
<The optimized loop>
**!oat$ install Exchange (Number of loop depth, Number of loop depth, …) region end**

The end-user specifies the OpenMP optimized loop with the above parameter <**Number of loop depth**> for the <**Exchange**> construct created by ppOpen-AT. Only one directive of OpenMP is accepted inside the optimized loop.

### B. Target Application and An Example

To explain the proposed method, we use the application software, Gyro Kinetic Vlasov (GKV) [13]. GKV is a program developed by T. Watanabe for use in the plasma simulation field.

The optimized loop is a quadruple loop. (See Figure 1). It appears in the **exb_realspcal** subroutine of the plasma turbulence analysis code in GKV [13]. This loop is transformed by the ppOpen-AT collapse function into two-nested loops shown in Figure 2 and Figure 3.

The proposed method can transform the loop in Fig. 1 into three loops, which are shown in Figure 4, Figure8, and Figure 10. In addition, the proposed method can be applied to loops transformed by the collapse function and can be further transformed into four loops. This code is shown in Fig. 4 to Fig. 10. For example, the following directive can generate code shown in Fig. 4 through Figure 7:

> **!oat$ install LoopFusion region start**
> **!oat$ install Exchange (1) region start**
> <Codes in Fig.1>
> **!oat$ install Exchange (1) region end**
> **!oat$ install LoopFusion region start**

The above directive <**LoopFusion**> is a loop collapse operation for the optimized loop specified by region start – region end. In this case, only one nested loop is specified, then code for all combinations of loop collapse for the optimized loop is generated (See Fig. 4 to Fig.7).

```
do iv = 1,  2*nv
!$OMP parallel do private(mx, my)
 do iz = (-nz),  nz-1
  do mx = ist_xw,  iend_xw
   do my = 0,  nyw
    ! calculation kernel
    wkdf1_xw(my,mx,iz,iv) =
     cmplx(real(wkdf1_xw(my,mx,iz,iv),kind=DP)  &
     * real(wkeyw_xw(my,mx,iz)-cs1 *vl(iv)  &
     * wkbyw_xw(my,mx,iz),kind=DP)-  &
      real(wkdf2_xw(my,mx,iz,iv),kind=DP)  &
     *real(wkexw_xw(my,mx,iz)-  &
      cs1*vl(iv)*wkbxw_xw(my,mx,iz),kind=DP ) &
     , aimag(wkdf1_xw(my,mx,iz,iv))  &
     *aimag(wkeyw_xw(my,mx,iz)-cs1*vl(iv)  &
     * wkbyw_xw(my,mx,iz))-  &
      aimag(wkdf2_xw(my,mx,iz,iv))  &
     * aimag(wkexw_xw(my,mx,iz)-  &
      cs1*vl(iv)*wkbxw_xw(my,mx,iz)) &
     , kind=DP ) * cef
  enddo enddo enddo
!$OMP end parallel do
enddo
```
Figure 1: Original quadruple loop. (The target of tuning.)

```
do iv = 1,  2*nv
!$OMP parallel do private(mx, my, mx_my)
 do iz = (-temp_nz),  temp_nz-1
  do mx_my = 1 ,  (iend_xw-ist_xw+1)*(nyw-0+1)
   mx=mod((mx_my-1)/(nyw-0+1),  (iend_xw-  &
    &ist_xw+1))+ist_xw
   my = mod((mx_my-1),  (nyw-0+1)) + 0
    ! calculation kernel
  enddo enddo
!$OMP end parallel do
Enddo
```
Figure 2: x and y-loop collapse created by ppOpen-AT. (xy collapse.)

```
do iv = 1,  2*nv
!$OMP parallel do private(mx, my, iz)
 do iz_mx_my = 1,  (temp_nz-1-(-temp_nz)+1)* &
   (iend_xw-ist_xw+1)*(nyw-0+1)
   iz = mod( (iz_mx_my-1)/((iend_xw-ist_xw+1)* &
    (nyw-0+1)),  (temp_nz-1- &
    (-temp_nz)+1)) + (-temp_nz)
   mx = mod((iz_mx_my-1)/(nyw-0+1),  &
    (iend_xw-ist_xw+1)) + ist_xw  &
   my = mod((iz_mx_my-1),  (nyw-0+1)) + 0
    ! calculation kernel
  enddo
!$OMP end parallel do
Enddo
```
Figure 3: z, x and y-loop collapse created by ppOpen-AT. (zxy collapse.)

```
!$OMP parallel do private(iz, mx, my)
do iv = 1,  2*nv
 do iz = (-nz),  nz-1
  do mx = ist_xw,  iend_xw
   do my = 0,  nyw
    ! calculation kernel
enddo enddo enddo enddo
!$OMP end parallel do
```
Figure 4: Directive to the outer-most loop.

```
!$OMP parallel do private(iz, mx, my, mx_my)
do iv = 1,  2*nv
 do iz = (-temp_nz),  temp_nz-1
  do mx_my = 1 ,  (iend_xw-ist_xw+1)*(nyw-0+1)
   mx=mod((mx_my-1)/(nyw-0+1),  (iend_xw- &
    ist_xw+1))+ist_xw  &
   my = mod((mx_my-1),  (nyw-0+1)) + 0
    ! calculation kernel
enddo enddo enddo
!$OMP end parallel do
```
Figure 5: Directive to the outer-most loop and x and y loop collapse. (xy collapse.)

```
!$OMP parallel do private(iz, mx, my, iz,
iz_mx_my)
do iv = 1,  2*nv
 do iz_mx_my = 1,  (temp_nz-1-(-temp_nz)+1)* &
```

```
    &(iend_xw-ist_xw+1)*(nyw-0+1)
  iz = mod( (iz_mx_my-1)/((iend_xw-ist_xw+1)* &
    (nyw-0+1)),  (temp_nz-1- &
    (-temp_nz)+1)) + (-temp_nz) &
  mx = mod((iz_mx_my-1)/(nyw-0+1),  &
    (iend_xw-ist_xw+1)) + ist_xw
  my = mod((iz_mx_my-1),  (nyw-0+1)) + 0
  ! calculation kernel
enddo enddo
!$OMP end parallel do
```

Figure 6: Directive to the outer-most loop and z, x, and y-loop collapse. (zxy collapse.)

```
!$OMP parallel do private(mx,my,iv,iz)
do iv_iz_mx_my = 1 , (2*nv)*(temp_nz-1-(-temp_nz)+1)*
&
    (iend_xw-ist_xw+1)*(nyw-0+1)
  iv = (iv_iz_mx_my-1)/((temp_nz-1-(-temp_nz)+1)* &
    (iend_xw-ist_xw+1)*(nyw-0+1)) + 1
  iz = mod((iv_iz_mx_my-1)/((iend_xw-ist_xw+1)*(nyw-
&
    0+1)),(temp_nz-1-(-temp_nz)+1)) + (-temp_nz)
  mx = mod((iv_iz_mx_my-1)/(nyw-0+1), &
    (iend_xw-ist_xw+1)) + ist_xw
  my = mod((iv_iz_mx_my-1),(nyw-0+1)) + 0
  ! calculation kernel
enddo
!$OMP end parallel do
```

Figure 7: Directive to the outer-most loop and v, z, x and y-loop collapse. (vzxy collapse.)

```
do iv = 1,  2*nv
  do iz = (-nz),  nz-1
!$OMP parallel do private(my)
    do mx = ist_xw,  iend_xw
      do my = 0,  nyw
        ! calculation kernel
      enddo enddo
!$OMP end parallel do
enddo enddo
```

Figure 8: Directive to the third loop from the outside.

```
do iv = 1,  2*nv
  do iz = (-temp_nz),  temp_nz-1
!$OMP parallel do private(mx, my)
    do mx_my = 1 ,  (iend_xw-ist_xw+1)*(nyw-0+1)
      mx=mod((mx_my-1)/(nyw-0+1), (iend_xw- &
        ist_xw+1))+ist_xw
      my = mod((mx_my-1),  (nyw-0+1)) + 0
      ! calculation kernel
    enddo
!$OMP end parallel do
enddo enddo
```

Figure 9: Directive to the second loop from the outside and x and y-loop collapse. (xy collapse)

```
do iv = 1,  2*nv
  do iz = (-nz),  nz-1
    do mx = ist_xw,  iend_xw
!$OMP parallel do private
      do my = 0,  nyw
        ! calculation kernel
      enddo enddo enddo
!$OMP end parallel do
enddo
```

Figure 10: Directive to the innermost loop.

## C. Experimental Results

In our experimental supercomputer environment, we used the Fujitsu PRIMEHPC FX100 (FX100), installed in the Information Technology Center, Nagoya University. The configuration of the FX100 is shown in Table I.

TABLE I.    SPECIFICATION OF THE FX100

| Contents | Specification | Details |
|---|---|---|
| CPU | SPARC64 XIfx (2.2 GHz) | *32 cores per node, 2 assistant cores. *2 sockets. Each socket has 16 cores. *Non-uniform Memory Access (NUMA) |
| Memory | 32 GB/node | Bandwidth: 480 GB/s (in 240GB/s, out 240 GB/s.). |
| Theoretical Peak | 1.1264 TFLOPS | |
| Cache Configuration | L1: 64 KB (Instruction/ Data separation, per core) L2: 24 MB (Shared with 32 cores) | Cache Organization: 4 Ways Set Associative. |
| Compiler | frtpx: Fujitsu Fortran Driver Version 2.0.0 P-id: T01815-01 | Options: -Kfast -Qt -Cpp -X9 -fs -fw -Kopenmp |

In this experiment, each loop from Fig. 1 to Fig. 10 was executed under the following conditions.
- Matrix sizes:
  - iv=1, 2*nv : From 1 to 16 (loop length: 16 )
  - iz=-nz, nz-1 : From -8 to 7 (loop length: 16 )
  - mx=ist_xw, iend_xw: From 0 to 127 (Loop length: 128 )
  - my=0, nyw : From 0 to 64 (Loop length: 65 )
- The number of threads in OpenMP is set to 32.

- 1000 iterations of the optimized loop.

The execution time was measured and a graph was created that shows the speedup using ppOpen-AT and the speedup using the proposed method with the original loop. The results are shown in Figure 11.

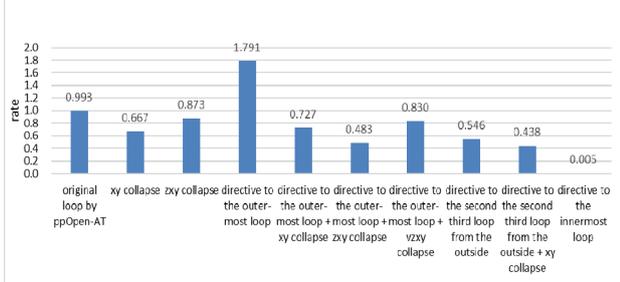

Figure 11: Speedup with the original loop in GKV.

According to Fig. 11, the execution with a changed loop position and the OpenMP directive to the outer-most loop was the fastest. It is 1.791 times faster than the execution of the original loop. This indicated that the AT function was critical for improved execution time in this application.

## IV. A NEW AT FUNCTION : DYNAMICE THREAD CHANGE IN OPENMP

### A. Proposed Method

In AT using ppOpen-AT, software developers set performance parameters. After releasing the software with AT function by ppOpen-AT, the values of performance parameters are changed by measuring the execution time for each tuning candidate when AT is performed in install-time or before execute-time. AT results for the values of performance parameters are referenced when the program is executed.

In this section, we propose a new AT function using the number of OpenMP threads as a performance parameter in addition to the AT function described in Section 3. Basically, we can implement the new function based on a conventional AT framework using ppOpen-AT, but dynamic change of the number of OpenMP threads is required. To accomplish this modification, the application programming interface (API) for OpenMP, named **omp_set_num_threads**, can be used to change the number of threads in OpenMP.

The proposed new AT function procedures are as follows.
1. User settings: End-user sets problem size, number of MPI processes, and maximum number of OpenMP threads.
2. AT execution: Perform AT for changing the number of threads for all candidates. Perform AT for the other performance parameters for all candidates. This AT timepoint is called "before execution" AT in FIBER [7]. Then find the best number of OpenMP threads for each candidate.
3. Perform large-scale execution:
   i. Call target routine.
   ii. Determine the best number of OpenMP threads using procedure 2.
   iii. Determine the best values of parameters using procedure 2.
   iv. Execute the target routine with the best number of threads and the best performance parameter values.

In code generated by ppOpen-AT, each tuning candidate is formed by a subroutine. Hence it is easy to change the number of OpenMP threads in the generated subroutine. For example, we use a maximum of 32 threads. We can change the thread number to *NumThread* and the generated code will automatically make the thread change by using **omp_set_num_threads**, such as:

Subroutine AutoGenCode1(….)
call **omp_set_num_threads** ( *NumThread* )
   <Auto-generated candidate code by ppOpen-AT>
call **omp_set_num_threads** ( 32 )
return
end

### B. Target Application and An Example

The target application is ppOpen-APPL/FDM, which is based on seismic simulation code Seism3D [14].

ppOpen-APPL/FDM is provided by the ppOpen-HPC project. The library supports hybrid execution of OpenMP and MPI. An AT function is implemented in ppOpen-APPL/FDM using the code selection function in ppOpen-AT. In this evaluation, we use the code generated by ppOpen-AT in ppOpen-APPL/FDM.

The experimental computer environment is the FX100. The configuration of the FX100 is the same as in Section 3.2. We used "-Kfast" and OpenMP as compilation options.

In this experiment, the target is the routine update_stress, which occupies 35% of the total execution time [13]. The experiment was performed by rewriting code for **OAT_InstallRoutines.f90**. The AT area is automatically generated by ppOpen-AT by specifying before-execution AT. As we mentioned, in before-execution AT, the computation kernel of update_stress is tuned by changing of the number of OpenMP threads by using **omp_set_num_threads**. After the update_stress, **omp_set_num_threads** computation, the original number of threads is restored.

The experiment was conducted under the following conditions.
- 8 nodes.

- 2000 time steps. The total time was the execution time with 2000 time steps.
- The number of MPI processes was eight.
- The thread change experiment was performed with the fastest number of threads from the before-execution AT.
- Execution without the thread change (conventional method) was set to the maximum number of OpenMP threads in a node for the FX100.

### C. Experimental Results

Figure 12 shows the increased execution speed for the target routine when using thread change at run-time as compared to the execution time without the thread change (conventional method). It is noted that total execution times for the whole program are considered in Fig. 12

### D. Discussion

Fig.12 indicates that in this experiment, by changing the number of threads, the execution time was improved by up to 1.003 times compared to the original loop using conventional methods (without changing the number of threads).

Although the improvement was a small number, it also shows that the overhead of changing the number of threads with **omp_set_num_threads** is also small. This means we can change the number of threads frequently at run-time using AT on the FX100. Therefore, the proposed AT method changing the number of threads can be useful for run-time AT suitable. It also indicates that increasing the number of AT candidates may be more effective because execution time of changing the number of threads is small.

## V. EVALUATION OF THE COMBINATION OF THE TWO PROPOSED METHODS

### A. Overview

We applied the AT functions outlined in Sections 3 and 4 simultaneously to evaluate the effect of these combined functions.

### B. Target Application and Example

The target application was the plasma turbulence analysis code, GKV, used in Section 3.2. The target subroutine was **exb_realspcal**, described in Section 3.2. The program was executed with the number of OpenMP threads changed at run-time.

### C. Experimental Results

Fig.12 indicates that in this experiment, by changing the number of threads, the execution time was improved by up to 1.003 times compared to the original loop using conventional methods (without changing the number of threads).

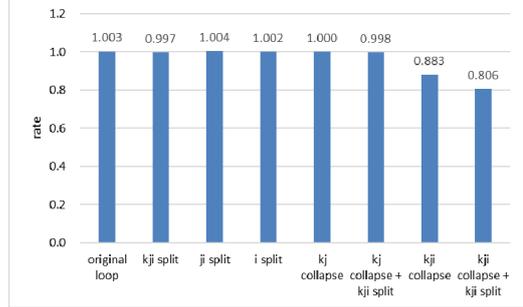

Figure 12: Overall increase in execution speed by changing the number of threads in Seism3D.

Figure 13 shows the execution rate for the original loop along with the two proposed methods. The rate in Figure 14 was computed based on the execution time in Fig. 13. The AT function for loop transformation in Fig. 13 and Fig. 14 was the same as in Section 3.3. Fig. 14 shows that the increase is obtained by changing the number of threads when the largest number of threads (32) is specified in the FX100.

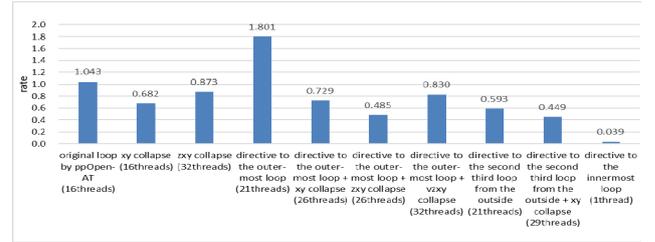

Figure 13: Increased execution speed as compared to the original loop in GKV with loop transformation and thread change. The numbers in the parentheses are the optimal number of OpenMP threads in each execution.

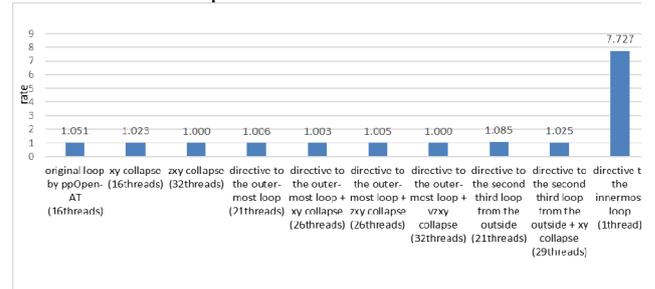

Figure 14: Increased execution speed compared to the maximum thread execution (32 threads) of each loop in GKV. The numbers in the parentheses are the optimal number of OpenMP threads in each execution.

### D. Discussion

According to the rate in directives to the outer-most loop in Fig. 13, there was an increase in speed of 1.801 times that of the execution for the original loop. It confirms an (1.791 times in Fig. 13) * (1.006 times in Fig.14) = (1.801 times in Fig.13) increase in execution from the original code.

In Fig. 14, we see an interesting change in the rate by applying directives to the innermost loop (1 thread), 7.727 times faster than the execution in the original loop collapsed code. On the other hand, the fastest loop in Fig. 14 applies directives to the outer-most loop. It performs 1.006 times faster than the unchanged thread (conventional method).

One of major reasons for the extreme increase in the directives to the inner-most loop execution is its small loop length. As we explained, the OpenMP optimized loop is the my-loop, which has a loop length of 65. Each thread has only a loop length of 2 when using a 32-thread execution for this kernel. This is too short to implement pipelining. This is one of the reasons to obtain a larger loop length (65) by using a 1-thread execution.

## VI. Related Work

Loop transformation techniques as a function of AT described in this paper are not a new idea [6][8-9][16-19]. ADAPT [17] proposes a framework for loop transformation with fusion of multiple splits to target computation kernels. Xevolver [6] can easily define loop transformation as a recipe on its framework. This definition for loop transformation is also called "Recipes for Code Generation." [18] ppOpen-AT [8][9] also provides loop collapse and loop split as AT functions.

The main contribution of this work is a new AT function that changes the number of threads dynamically with loops automatically transformed by the AT system. It also is easily implemented as an AT function in ppOpen-AT using an OpenMP API. In this paper we describe the effectiveness of the new AT function with several applications—GKV and ppOpen-APPL/FDM.

## VII. Conclusion

In this research, to achieve high performance for numerical calculation software, we proposed two new AT functions for diversified computer architectures. The AT functions are not currently implemented functions in ppOpen-AT.

The first AT function is a loop transformation that changes the optimized loop (position of directives) in an OpenMP. This change can be adapted for loop transformation, such as loop collapse, using ppOpen-AT.

The second AT function is the dynamic change of the number of OpenMP threads. Conventional execution uses a fixed number of threads. By adapting the new AT function, the number of threads is changed in each optimized loop at run-time. After execution of each optimized loop, the number of threads is set to the maximum number of threads that is specified by the end-user.

We also evaluated the efficiency of the two AT functions. Results indicated that a maximum increase in execution speed of 1.801 times that of the original loop was obtained as compared to conventional execution, using AT functions in GKV as a benchmark. This indicates that proposed AT function can easily apply typical scientific simulation codes, such as plasma simulation.

As a future work, evaluation of these two new AT functions with other loops in several benchmarks or application software is necessary.

### Acknowledgment

This work was supported by JSPS KAKENHI Grant Number JP18K19782 and JP19H05662. The authors would like to thank to Professor Tomo-Hiko Watanabe and Dr. Shinya Maeyama for providing us knowledge of the GKV code.